\documentclass[%
 reprint,
 amsmath,amssymb,
 aps,
prl,
]{revtex4-2}

\usepackage{graphicx}% Include figure files
\usepackage{dcolumn}% Align table columns on decimal point
\usepackage{bm}% bold math
\usepackage{color}
\usepackage[normalem]{ulem}
\begin{document}

\preprint{APS/123-QED}

\title{Morphological instability at topological defects\\ 
in a three-dimensional vertex model for spherical epithelia}

\author{Oliver M. Drozdowski}
\affiliation{Institute for Theoretical Physics and BioQuant, Heidelberg University, 69120 Heidelberg, Germany}
%\affiliation{Max Planck School Matter to Life}
\author{Ulrich S. Schwarz}%
 \email{Corresponding author: schwarz@thphys.uni-heidelberg.de}
\affiliation{Institute for Theoretical Physics and BioQuant, Heidelberg University, 69120 Heidelberg, Germany}

\date{\today}% It is always \today, today,
             %  but any date may be explicitly specified

\begin{abstract}
Epithelial monolayers are a central building block of complex organisms. Topological defects have emerged as important elements for single cell behavior in {flat} epithelia. {Here we theoretically study such defects in a three-dimensional vertex model for spherical epithelia like cysts or intestinal organoids. 
We find that they lead to the same generic morphological instability to an icosahedral shape as it is known from spherical elastic shells like virus capsids, polymerized vesicles or buckyballs. 
We derive analytical expressions for the effective stretching and bending moduli as a function of the
parameters of the vertex model, in excellent agreement with computer simulations. These equations accurately predict both the buckling of a flat epithelial monolayer under uniaxial compression and the faceting transition around the topological defects in spherical epithelia. We further show that localized apico-basal tension asymmetries allow them to reduce the transition threshold to small system sizes.}
 \end{abstract}

%\keywords{Suggested keywords}%Use showkeys class option if keyword
                              %display desired
\maketitle

%\tableofcontents
{\it Introduction.}
Epithelial monolayers are a central element of
the architecture of complex organisms. They
separate different compartments, can form
highly convoluted shapes and have exceptional mechanical
properties. In particular, they can quickly
undergo transitions between fluid-like and elastic properties 
\cite{PerezGonzalez_Trepat_NatCellBiol2021_2d_intestinal_organoid,Khalilgharibi_Charras_NatPhys19_Epithelial_stretching_viscoelasticity}, driven e.g.\ by cell density or
the aspect ratios of single cells. In general, the properties of the
single cells are essential to understand
the physical properties of epithelial 
monolayers. Topological defects defined
by the neighborhood relations of the single cells
have emerged as especially important elements
for transformations in epithelial monolayers 
\cite{Shankar_NatRevPhys_2022_Review_topological_active_matter}. For example, it has been found that single cells
tend to be extruded at such topological defects 
\cite{Saw_Nat_2017_topological_defects_extrusion}.
While topological defects are a natural
ingredient of hydrodynamic theories 
\cite{Hoffmann_Giomi_SciAdv_22_Defect_mediated_morphogenesis},
it is challenging to include them in elastic 
descriptions of monolayers \cite{Hannezo_PNAS14_Epithelial_morphology_theory, Latorre_Nat18_Epithelial_domes, Ziherl_PRL23_Wrinkling_1d_VM}.

Here we show that spherical epithelia like cysts or intestinal organoids
are a natural starting point to study the global effects of topological defects
in epithelial monolayers. They are experimentally very accessible
and of large biomedical relevance \cite{Clevers_Cell2016_Review_organoids}.
Due to Euler's polyhedron theorem, they necessarily have to include
twelve pentagons in the close-packed
tiling of the spherical surface \cite{Lidmar_PRE_2003_icosahedral_instability_shells}.
In order to combine these topological defects
with the typical mechanical properties
of epithelial monolayers, we employ 
a three-dimensional (3D) vertex model (VM), in which 
cells are described as polyhedra with a fixed volume and with polygonal faces 
contributing to the total energy through surface tensions \cite{Krajnc_PRE18_VM_fluidization}. 
The 3D VM has been used before for modeling spherical epithelia \cite{Rozman_NatComms2020_organoid_3d_VM, Hannezo_PNAS14_Epithelial_morphology_theory, Eiraku_Nat11_Optic_cup_organoids_VM, Okuda_SviAdv18_Mechanical_feedback_optic_cup_VM,Yang_Cell21_Intestinal_organoid_VM}, 
{but coarse-graining procedures have not been able to fully address 
the role of topological defects in such a setting.}

\begin{figure}[!t]
	\centering
	\includegraphics[width=\columnwidth]{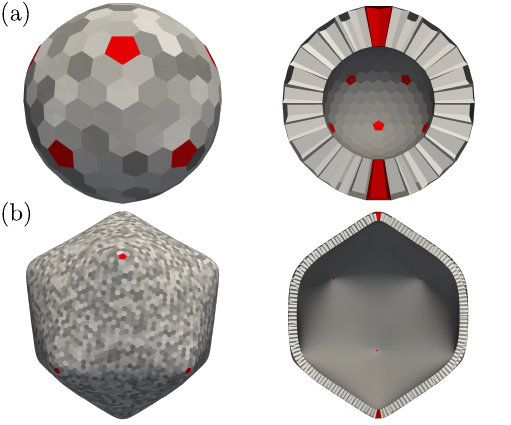}
	\caption{{Spherical epithelia described by a 3D vertex model experience an icosahedral instability
 that is well known for spherical elastic shells like virus capsids, polymerized vesicles or buckyballs. 
 (a) A small epithelial shell stays spherical. (b) A large epithelial shell becomes faceted with icosahedral symmetry.}
 Cells are arbitrarily colored in grey and pentagonal cells, {which are topological defects}, are shown in red.}
\label{fig:shells}
\end{figure}

By simulating the 3D VM for complete spherical epithelial shells, we discovered an icosahedral faceting instability.
While small shells with few cells have a spherical shape (Fig.~\ref{fig:shells}(a)), larger icosahedral shells experience 
conical instabilities at the pentagonal defects (Fig.~\ref{fig:shells}(b))
\footnote{{Here and in the following
we use the Caspar-Klug classification for virus 
capsids to indicate shell size. 
The Caspar-Klug indices $(h,k)$ 
describe the steps in the hexagonal lattice to go from one pentagonal cell to another with respect 
to lattice vectors with an opening angle of $\pi/3$.
The two shells shown in Fig.~\ref{fig:shells}(a) and (b) have sizes $(5.0)$ and $(19,12)$, respectively}}.
This transition is well known for two-dimensional elastic crystals
\cite{Seung_88_PRL_Conical_instability_disclination,Lidmar_PRE_2003_icosahedral_instability_shells},
including virus capsids \cite{Lidmar_PRE_2003_icosahedral_instability_shells, Nguyen_PRE05_viral_capsid_spontaneous_curvature},
polymerized vesicles \cite{blaurock_small_1979_vesicles_faceting,Lin_SoftMatter2011_block_copolymer_vesicle_faceting}, and buckyballs \cite{Witten_EuPhysLett93_Fullerene_Ball_icosahedral_instability},
{but has not been described before for spherical epithelia.} 
Our numerical results suggest
that a continuum limit exists for the 
3D VM that like thin elastic shells
contains both stretching and bending energies.
{Here we show how such a coarse-graining procedure
can be performed and that it explains the morphological instability.
We further show that the threshold for this instability can be 
actively controlled by epithelia through apico-basal polarity.}

{\it Continuum model.}
We start with the Hamiltonian of a 3D VM with apical, basal and lateral faces,
\begin{equation}\label{eq:hamiltonian_vertex}
    E_\mathrm{VM}=\sum_\mathrm{cells} \left( \Gamma_\mathrm{a} A_\mathrm{a} + \Gamma_\mathrm{b} A_\mathrm{b} + \frac{1}{2} \sum_\text{lateral faces} \Gamma_\mathrm{l} A_\mathrm{l} \right),
\end{equation}
with surface tensions $\Gamma_i$ and areas $A_i$ for apical,
basal and lateral faces, respectively ($i=\mathrm{a},\mathrm{b},\mathrm{l}$).
The factor $1/2$ avoids membrane double counting.
Assuming volume $V$ being conserved, we non-dimensionalize energy with 
$\Gamma_\mathrm{l}$ and length with $V^{1/3}$.

To derive a thin-plate theory from the 3D VM, 
we consider the nonlinear theory of moderately bent plates, 
where the total energy is given by stretching and bending energy contributions \cite{Landau_Lifschitz}. For the in-plane stretching energy 
we assume the usual energy density
\begin{equation}\label{eq:stretching_energy_density}
    e_\mathrm{stretch} = \frac{1}{2} \left( 2\mu \varepsilon_{ij}^2 + \lambda \varepsilon_{kk}^2\right), 
\end{equation}
with two-dimensional strain tensor $\bm{\varepsilon}$ and 
Lam\'e coefficients $\mu$ and $\lambda$.
We determine the Lam\'e coefficients 
in a flat configuration in a mean-field fashion,
following ideas from earlier work on 2D VMs \cite{Staple_EPJE2010_mechanics_cell_packing_epithelia_VM, Moshe_PRL18_Geometric_Frustration_VM, Staddon_SM23_Non_affine_deformations_VM}. 
We consider a constant strain tensor for an individual cell, i.e.\ we assume strain to vary on a larger length scale than cell size. We assume a regular $n$-gonal lattice structure, which we will take to be a hexagonal lattice with $n=6$. 
We then determine the energy density for a given strain.
The equilibrium height can be obtained via minimization of Eq.~(\ref{eq:hamiltonian_vertex}) at constant volume.
For the lateral faces we employ an angular averaging method (described in detail in the supplement 
\footnote{See Supplemental Material [URL WILL BE ADDED AFTER PUBLICATION], 
    which includes details on the derivation of the stretching parameters 
    (I), including a nonlinear treatment (II), and of the bending parameters (III), 
    on the numerical implementation of the 3D VM (IV), on the finite element 
    calculation of the stretching energy 
    in the spherically bent epithelium (V),
    and on the post-buckling approximation for a compressed sheet (VI). 
    It also includes Refs.~\cite{CGAL, Brakke_92_Surface_Evolver}.}).
In addition we consider non-affine deformations,
as previously described for 2D models \cite{Staddon_SM23_Non_affine_deformations_VM, Murisic_BiophysJ15_2d_VM_Bending}. 
Non-affine relaxations correspond to additional 
relative deformations of the two sublattices that make up the hexagonal lattice  and 
can be included by allowing for an additional degree of freedom in the mid-plane 
shape, which allows for force balance at triple membrane junctions via angular 
relaxation.
Note that the $\pi/3$ rotational symmetry implies that our 2D continuum model has only
two elastic constants, exactly like an isotropic 2D material.

Considering the deformed areas and Taylor-expanding in the principal strains, we find $2\mu=\lambda=\Gamma_\mathrm{a}+\Gamma_\mathrm{b}$, or, equivalently,
\begin{equation} \label{eq:elastic_constants}
    Y = \frac{4\mu(\mu+\lambda)}{2\mu+\lambda} = \frac{3}{2}(\Gamma_\mathrm{a}+\Gamma_\mathrm{b}),\quad \nu=\frac{\lambda}{2\mu+\lambda}=\frac{1}{2},
\end{equation}
for the two-dimensional Young's modulus and Poisson ratio, respectively. The Young's modulus does not
depend on the lateral tension $\Gamma_\mathrm{l}$, because
both $\Gamma_\mathrm{a/b}$ and Y are in units of $\Gamma_\mathrm{l}$. 
The reason is that changes in $\Gamma_\mathrm{l}$ will affect height and edge length in such a manner that the energy density 
stays the same. 
%Interestingly, the numerical factor $3/2$ in Eq.~(\ref{eq:elastic_constants}) 
%results from the effect that the lateral areas will change due to volume conservation. 
A 2D Poisson ratio of $1/2$ means that the sheet is compressible 
(incompressible materials in 2D have $\nu = 1$), because it can exchange
material between the in-plane and out-of-plane dimensions.
{In addition, we formulated the stretching energy in the fully nonlinear setting (see supplement \cite{Note2}).
The resulting energy density does not match standard hyperelastic models \cite{Howell_Ockendon_Solid_Mechanics}. Thus in the following
we restrict ourselves to the first (cubic) correction
to the linear theory as obtained by a Taylor expansion. 
This yields $Y_\text{nl}=\left(\frac{3}{2}-\frac{7}{4}\varepsilon_{xx}\right)(\Gamma_\mathrm{a}+\Gamma_\mathrm{b})$ and $B_\text{nl}=\left(\frac{3}{2}-\varepsilon_{xx}\right)(\Gamma_\mathrm{a}+\Gamma_\mathrm{b})$ for the Young's modulus and bulk modulus in uniaxial and isotropic stretching, respectively.}

For the bending energy density we assume the Helfrich form 
\begin{equation}\label{eq:bending_energy_density}
    e_\mathrm{bending} = \frac{\kappa}{2} (H-c_0)^2 + \kappa_\mathrm{G} K,
\end{equation}
with bending rigidity $\kappa$, mean total curvature $H=c+c'$ with 
the principal curvatures $c$ and $c'$, spontaneous curvature $c_0$, 
saddle splay modulus $\kappa_\mathrm{G}$ and Gauss curvature $K=cc'$.
Rozman et al. \cite{Rozman_NatComms2020_organoid_3d_VM} have proposed a 
method to determine these quantities for the 3D VM using quadratic lattices.
We generalized this to hexagonal lattices and adapted it such that we can 
formulate a theory for moderate bending.

Consider a cell which is bent with principal curvatures $c$ and $c'$ and with unchanged center height. 
Volume conservation and curvatures determine the apical, basal and lateral face 
areas after deformation. 
Normalizing to the undeformed mid-plane area (for consistency and different from Ref.~\cite{Rozman_NatComms2020_organoid_3d_VM}) 
and Taylor-expanding with respect to $ch$ and $c'h$ yields the bending energy density 
(see supplement for full derivation \cite{Note2}).
Like for the stretching part, we also consider
non-affine deformations, which leads to
a correction factor $k_c$. For $n=6$ we find
\begin{equation}\label{eq:bending_parameters}
\begin{aligned}
    \kappa &= \frac{9}{8}\, \frac{k_c}{ 2^{1/3}\, 3^{4/3}} (\Gamma_\mathrm{a} + \Gamma_\mathrm{b})^{1/3},\\
    \kappa_\mathrm{G} &= \left[ \frac{(\Gamma_\mathrm{a}+\Gamma_\mathrm{b})^2}{2} + \frac{3}{2} - \frac{9}{4} k_c \right] \frac{(\Gamma_\mathrm{a} + \Gamma_\mathrm{b})^{1/3}}{2^{1/3} 3^{4/3}}, \\
    c_0 &= \frac{4}{3}\, 2^{2/3}\, 3^{1/6}\, (\Gamma_\mathrm{a}+\Gamma_\mathrm{b})^{1/3}\, (\Gamma_\mathrm{b}-\Gamma_\mathrm{a}).
\end{aligned}
\end{equation}
Bending with $c\neq c'$ is accompanied by non-isotropic 
stretching in the apical and basal planes and angular relaxation,
similarly as in non-isotropic stretching of the mid-plane.
For $c=c'$ such non-affine relaxations do not occur since lateral membrane angles do not change. 
In this case the energy is identical to the case of $k_c=1$,
but otherwise $k_\mathrm{c}$ is a numerical factor that we obtain from
fitting to the simulation results for cylindrical surfaces.

\begin{figure}[!t]
	\centering
	\includegraphics[width=\columnwidth]{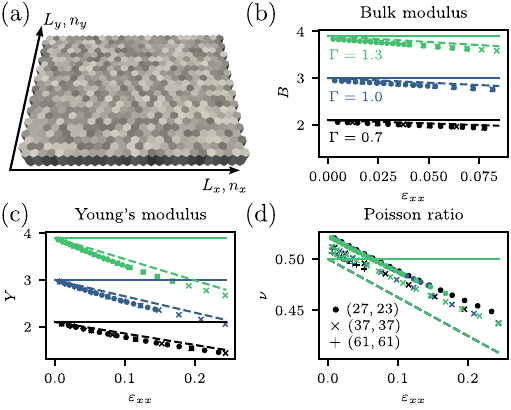}
	\caption{Stretching of a rectangular 3D VM monolayer. (a) Depiction of the simulated monolayer with arbitrary coloring of individual cells. (b) Bulk modulus $B$ for isotropic stretching with strain $\varepsilon_{xx}$. (c,d) Young's modulus $Y$ and Poisson ratio $\nu$ for uniaxial stretching with strain $\varepsilon_{xx}$. Colors indicate different surface tensions $\Gamma$ and symbols different plate sizes $(n_x, n_y)$. Dashed and solid lines 
    are the predictions from nonlinear and linear continuum theories, respectively.}
\label{fig:stretching}
\end{figure}

The results from Eq.~(\ref{eq:bending_parameters}) differ in several essential
ways from the known formulae for thin elastic plates. 
For a thin elastic plate the bending rigidity would scale like $\kappa\propto Yh^2$ 
(with 2D Young's modulus $Y=Y_\mathrm{3D}h$).
With our result for $Y$ and the reference height $h\propto(\Gamma_\mathrm{a}+\Gamma_\mathrm{b})^{2/3}$ (compare supplement \cite{Note2}), this would lead to $\kappa\propto(\Gamma_\mathrm{a}+\Gamma_\mathrm{b})^{7/3}$.
As seen in Eq.~(\ref{eq:bending_parameters}), the bending rigidity depends more weakly on the tensions, because we do 
not have to consider the area changes along the entire height.
In fact, compression on one of the sides will lower the 
energy instead of increasing it, as it is the case in 3D elastic plates.
The dependence is much stronger, however, for the saddle-splay modulus, as we cannot compensate 
for area changes of the polygonal faces
via shape changes (e.g. from rectangles to trapezoids) when both curvatures do not vanish.
We have the same leading order scaling as we would have for elastic plates 
because here the apical and basal area 
changes for a given curvature also enter quadratically in the cell height.
The spontaneous curvature depends on the apico-basal tension asymmetry, 
as this introduces a preferred curvature to minimize the energies. For
$\Gamma_\mathrm{a}=\Gamma_\mathrm{b}$ it vanishes as expected.

Like for a thin plate, the full deformation energy can now be calculated as
\begin{equation}\label{eq:total_energy_bending_stretching}
    E = \int (e_\mathrm{stretch} + e_\mathrm{bending})\, \mathrm{d}S.
\end{equation}
For moderately bent plates there is an additional
coupling between the two terms. 
A mid-plane deflection $f(x,y)$ will contribute to the strain tensor as
$\varepsilon_{ij} = \left( \partial_i u_j + \partial_j u_i + \partial_i f \partial_j f \right)/2$,
with deformation $u_i$ in $i$-direction.

{\it Stretching and bending of a flat sheet.}
To test the continuum theory by computer simulations, 
we have implemented the 3D VM, Eq.~(\ref{eq:hamiltonian_vertex}), as module in the
software suite Chaste \cite{Chaste_Software_2020}, 
similarly to Ref.~\cite{Krajnc_PRE18_VM_fluidization}
(details in the supplement \cite{Note2}).
For stretching we implemented a finite-size rectangular plate of hexagonal cells.
The monolayer consists of $n_x$ and $n_y$ cells in the $x$ and $y$-directions, respectively (Fig.~\ref{fig:stretching}(a)).
For %the remainder of this work 
now we assume $\Gamma_\mathrm{a}=\Gamma_\mathrm{b}=\Gamma$, i.e.\ we do not consider a spontaneous curvature $c_0$ from apico-basal polarity.

First we considered isotropic stretching with edge stresses $\sigma_{xx}=\sigma_{yy}$ and measured the 
effective bulk modulus as
$B = \lambda+\mu = \varepsilon_{xx}/2\sigma_{xx}$. Then
we considered uniaxial stretching with edge
stresses $\sigma_{yy}=0$ and 
measured Young's modulus as $Y=\sigma_{xx}/\varepsilon_{xx}$ and Poisson ratio as $\nu=-\varepsilon_{yy}/\varepsilon_{xx}$. 
Fig.~\ref{fig:stretching}(b-c) demonstrates excellent
agreement between the simulations results (symbols)
and the nonlinear continuum theory (dashed lines). Moreover
the elastic moduli $B$ and $Y$ from Eq.~(\ref{eq:elastic_constants}) (solid lines) correspond
exactly to the limiting cases of vanishing strain.
The Poisson ratio $\nu$ is close to $1/2$ as
predicted by Eq.~(\ref{eq:elastic_constants}). In the following, we will 
mainly discuss the case of linear elasticity,
but will come back to our nonlinear results when needed.

\begin{figure}[!t]
	\centering
	\includegraphics[width=\columnwidth]{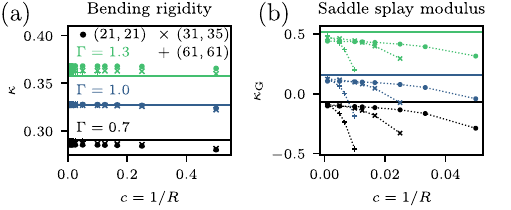}
	\caption{Bending of a rectangular 3D VM monolayer. (a) Bending rigidity $\kappa$ as determined from a cylindrically bent monolayer. (b) Saddle splay modulus $\kappa_\mathrm{G}$ as determined from a spherically bent monolayer.
 %\sout{(c) Cell height for a spherically bent rectangular monolayer with $\Gamma=1.3$, size $(31,35)$ and $c=1/20$ from finite element energy minimization of the continuum theory. (d) The same
 %   for the 3D VM in the undeformed mid-plane.}
 }
\label{fig:bending}
\end{figure}

Next we simulated bending of cylindrical and spherical surfaces.
Fig.~\ref{fig:bending}(a) shows the resulting bending rigidity $\kappa$ as function of curvature $c=1/R$ of a cylinder segment with mid-plane radius $R$. 
For $\Gamma=1$, we determine the non-affinity correction factor
by a fit as $k_c\approx 1.26$ and show that it is
caused by non-affine apical and basal deformations (compare
supplement \cite{Note2}). 
Our simulation results (symbols)
agree well with the prediction from continuum theory,  Eq.~(\ref{eq:bending_parameters}) (solid lines). Fig.~\ref{fig:bending}(b) shows the results
for the saddle splay modulus $\kappa_\mathrm{G}$. 
{For this stretching contributions, arising from the non-developable spherical deformation, were determined via finite element simulations, implemented with FEniCS \cite{Note2, Alnaes_2015_FEniCS}.}
Again we find good agreement with the theoretical
pediction from Eq.~(\ref{eq:bending_parameters}) (solid lines) 
for small curvature. The deviations
at larger curvature are related to finite-size effects,
including overestimation of the stretching energy
for differently sized plates. 

%Non-affine apical and basal deformations indeed justify the correction $k_c$ in the case $c\neq c'$. 
%In Fig.~\ref{fig:bending}(c) we show cylindrically bent cells with nodes projected onto the surface before and after energy minimization.
%We note that in both case the apical/basal polygons are not coplanar, 
%different from our continuum assumptions. In general, we
%find similar apical and basal deformations as in uniaxial mid-plane stretching.
%The spherical surface is not developable and as such stretching contributions cannot be neglected for large curvatures.
%To find the stretching energy we considered a spherical deflection %$f(x,y)=R-\sqrt{R^2-x^2-y^2}$ in the strain
%and minimized the corresponding stretching energy, cf.~Eq.~(\ref{eq:stretching_energy_density}), 
%using the finite element method, implemented with FEniCS \cite{Alnaes_2015_FEniCS}. 
%{We found good agreement between the simulations and the 
%predictions of the continuum theory for the height distribution} (compare supplement \cite{Note2}).

{\it Buckling of a compressed sheet.}
Our continuum theory effectively describes the epithelial monolayer as a moderately bent plate. 
A classical application of such a theory is plate buckling upon in-plane compression, 
which has also been demonstrated experimentally for epithelial monolayers \cite{Wyatt_NatMat20_Buckled_suspended_monolayer}, depending on both stretching and bending.
%For a simply-supported rectangular plate, compressed along the edges parallel to the $y$-direction with compressive stress $\sigma_{xx}$,
%the critical buckling stress is given by \cite{Timoshenko1961}
%\begin{equation}\label{eq:critical_buckling_stress}
%\sigma_\mathrm{crit} = \frac{\pi^2 \kappa}{L_x^2} \left( m + \frac{1}{m}\, \frac{L_x^2}{L_y^2}\right)^2
%\end{equation}
%where the multiplicity $m$ describes the number of half-waves along the $x$-direction and it is chosen such that the critical stress is minimal, e.g.\ $m=1$ for quadratic plates. We now use plate buckling 
%as another test of our continuum theory. 

\begin{figure}[!tb]
	\centering
	\includegraphics[width=\columnwidth]{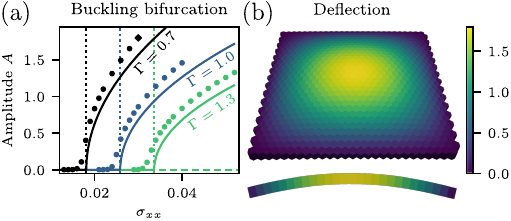}
	\caption{Buckling instability in a simply supported 3D VM sheet of size $(27,31)$ under uniaxial compression. (a) Buckling bifurcations in amplitude with the total compressive stress $\sigma_{xx}$ as control parameter for different tensions $\Gamma$. The solid lines are the continuum mechanics results for plate buckling, the dashed line is the unstable unbuckled state, the dotted line is the critical buckling stress, and the symbols are 3D VM simulations. (b) 3D VM simulation of deflected monolayer for $\Gamma=0.7$ and $\sigma_{xx}=0.03$, indicated in (a) as diamond marker. Below the 3D image a cross-sectional view is shown. Color indicates the mid-plane deflection of cell.}
\label{fig:buckling}
\end{figure}	

Fig.~\ref{fig:buckling}(a) shows the amplitudes of a simply-supported rectangular plate, compressed along the edges parallel to the $y$-direction with compressive stress $\sigma_{xx}$.
We assumed straight (but movable) edges, as if the plate was situated in a movable rigid frame, and found a bifurcation toward a bent state with one half-wave along both axes. 
The critical stress in the 3D VM is slightly smaller than the continuum expectation, which can be explained by the nonlinearities, which we have neglected in the mean-field model, and by finite-size effects of the plate.
The post-buckling amplitude for straight edges can be approximated within thin-plate theory \cite{Timoshenko1961} and 
is shown with solid lines.
For this the leading-order Fourier modes are considered and the energy is minimized for these modes (compare supplement \cite{Note2}).
We do see good agreement between this approximation and the 3D VM simulation results.
In Fig.~\ref{fig:buckling}(b) the mid-plane deflection is shown.
For large deflections we see a flattening of the profile with stronger deviations from a sinusoidal leading-order approximation, consistent with real thin elastic plates \cite{Rhodes_03_Post_buckling_thin_plates}.

{\it Topological defects and icosahedral instability.}
{The elastic framework derived above for epithelial monolayers suggests to
study the effect of topological defects on spherical shells in the same manner as
usually done for 2D elastic crystals. In our context, the disclinations
are the pentagonal cells in the hexagonal monolayer.}
For such a defect with disclinicity $s=2\pi/6$, the in-plane azimuthal 
stretching energy of a disc scales quadratically with the radius.
This deformation will become 
unstable toward a conical bending deformation with a logarithmic scaling in the energies for large radii \cite{Seung_88_PRL_Conical_instability_disclination, Lidmar_PRE_2003_icosahedral_instability_shells}.
Thus elastic shells of sufficient size, like
large virus capsids or buckyballs, undergo a shape
instability, in which each of the 12 pentagonal defects
becomes the corner of an icosahedron.

\begin{figure}[!t]
	\centering
	\includegraphics[width=\columnwidth]{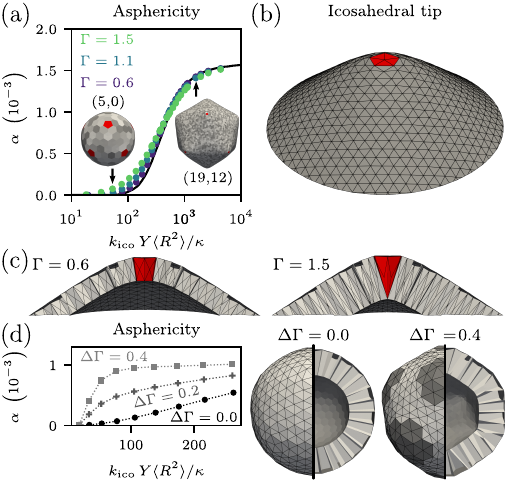}
	\caption{Icosahedral instability of 3D VM shells. (a) Asphericity {$\alpha=\langle(R-\langle R \rangle)^2/\langle R^2\rangle$} of the shells for different apical/basal surface tensions $\Gamma$ as function of the average cell-radius. 
 The radii are scaled by the ratio of the Young's modulus $Y$ and bending rigidity $\kappa$ with a nonlinearity correction $k_\mathrm{ico}$.  The solid curve is the continuum prediction \cite{Lidmar_PRE_2003_icosahedral_instability_shells}. 
 (b) A cut-out of the icosahedral tip for Caspar-Klug indices $(23,0)$ \cite{Note1} shows a conical deformation. 
 (c) For large tensions the cell height increases and for large radii the inner membrane collapses at the defect.
 {(d) Apico-basal tension asymmetry $\Delta\Gamma$ in the defects and their nearest neighbors lowers the buckling threshold. 
 Cells with $\Delta\Gamma\neq0$ are shaded in dark (size $(5,0)$, $\Gamma=1.1$).}}
\label{fig:icosahedral}
\end{figure}

As already shown in Fig.~\ref{fig:shells}, our computer
simulations of the 3D VM demonstrate exactly this scenario. 
To provide more details, Fig.~\ref{fig:icosahedral}(a) shows the 
asphericity $\alpha = \langle(R-\langle R \rangle)^2/\langle R^2\rangle$ of the cell centers as a function 
of the rescaled quadratic radius.
%{As common for viral capsids, we use the Caspar-Klug indices $(h,k)$ to indicate shell size.}
It is well known that the transition 
depends on the ratio of bending rigidity and Young's modulus,
which sets the relevant length scale $\sqrt{\kappa/Y}$ \cite{Seung_88_PRL_Conical_instability_disclination}.
In addition, we introduce a nonlinearity correction $k_\mathrm{ico}$. 
For an $s=2\pi/6$ disclinicity the azimuthal strain is as large as $0.2$ and we are in the strongly nonlinear regime, cf.~Fig.~\ref{fig:stretching}(c).
We find that a correction factor of $k_\mathrm{ico}\approx 1/2$ is necessary to account for this. With this scaling all curves
in Fig.~\ref{fig:icosahedral}(a) collapse onto the continuum limit (solid
line taken from the literature \cite{Lidmar_PRE_2003_icosahedral_instability_shells}) except for small radii.
This deviation can be understood to be a finite size effect as small radii correspond to few cells and large lattice constants compared to the radius.
Notice that we indeed find a conical deformation at the pentagonal tips of the icosahedron (Fig.~\ref{fig:icosahedral}(b)),
where the inner membrane can even collapse for large $\Gamma$ and thus large heights (Fig.~\ref{fig:icosahedral}(c)).
Experiments suggest $Y=200$~kPa\,$\mu$m and $V^{1/3}\approx10$~$\mu$m \cite{Charras_PNAS12_cell_stretching},
resulting in $\kappa\approx2200$~kPa\,$\mu\text{m}^3$ for the VM with $\Gamma=1$. 
The buckling threshold is known to be 
$k_\mathrm{ico}YR_\mathrm{crit}^2/\kappa\approx 154$ \cite{Lidmar_PRE_2003_icosahedral_instability_shells}.
The critical radius is thus $R_\mathrm{crit}\approx 60\,\mu\text{m}$, 
which is roughly the size at which 
intestinal organoids undergo budding \cite{Yang_Cell21_Intestinal_organoid_VM, Hartl_DevBio19_Organoids_crypts_blebbistatin}.

{In passive elastic shells, topological defects tend to form additional structures
such as defect scars, which screen the effect of the single defects \cite{bausch2003grain,GarciaAguilar_Giomi_PRE20_dislocation_screening}.
For flat epithelial monolayers it has already been established that
active processes modulate their elastic behavior \cite{Krajnc_PRE18_VM_fluidization}, thus also affecting the role of defects.
For spherical epithelia active, apico-basally polarized forces become essential for structure formation, as observed experimentally.
For example, in cell extrusion cells are pushed outward through contraction \cite{Rosenblatt_CurrBiol_2001_Extrusion_through_contraction,Fadul_CurrOpCellBio_2018_Forces_fates_extruding_cells},
and in budding organoids luminal (apical) contraction in buds facilitates curvature generation \cite{Yang_Cell21_Intestinal_organoid_VM}. 
To study such processes in our context, we add polarity in 
the pentagonal defect cells and their nearest neighbors, by using finite
values for $\Delta\Gamma = \Gamma_\mathrm{a}-\Gamma_\mathrm{b}$.
Fig.~\ref{fig:icosahedral}(d) shows that such concentration of curvature generation 
around the topological defects facilitates buckling at smaller radii, allowing for 
active control of the instability threshold. 
Thus the instability can already occur in a neighbourhood of a few hexagonal cells, with potential implications 
for organoid formation and cell extrusion in less structured epithelia.
Indeed, such hexagonal regions have been observed experimentally for epithelia with and without curvature \cite{Tang_NatPhys_2022_Hexagonal_islands_curved_epithelia, Armengol-Collado_NatPhys2023_Hexanematic_crossover_epithelia}.}

{In summary, here we have shown with computer simulations and analytical calculations
that with growing size, spherical epithelia should undergo the same morphological instability 
at topological defects that is known also for elastic shells such as virus capsids.
Our theory applies as long as the system is sufficiently regular 
and does not become too heterogeneous (e.g. by cell differentiation) before
the threshold is reached. Therefore we expect that experimentally
it might be observed best for highly regular epithelia, 
such as the retinal pigmented epithelium. Indeed, our theory might explain
the formation of drusen, which are spherical or conical out-of-plane deformations in the retina linked to makular degeneration \cite{Schlanitz_2019_ophthalmology_drusen_formation_retina, Ishibashi_Opthalmology86_drusen_em_budding, Usui_JCellSci18_Retinal_organoid_drusen}.}

{In the future, it has to be seen
how the elastic effects described here will be modulated by the dynamics
of epithelial monolayers, both on the level of single cells
\cite{Janshoff_BioChem21_Review_viscoelasticity_epithelial_cells} and on the tissue
level \cite{Bi_NatPhys15_Rigidity_transition_VM, Hernandez_PRE22_Anomalous_elasticity_rigidity_transition_VM}.
At any rate, however, our theory demonstrates that topological defects are not
only important for single cell behaviour, but have a strong
effect on the global properties of epithelial monolayers and thus could
mediate long-ranged effects.}

This research was conducted within the Max Planck School Matter to Life supported by the German Federal Ministry of Education and Research (BMBF) in collaboration with the Max Planck Society.
USS is a member of the clusters of excellence Structures (EXC 2181/1-390900948) and 3DMM2O (EXC 2082/1-390761711) funded by the Deutsche Forschungsgemeinschaft (DFG, German Research Foundation) under Germany’s Excellence Strategy as well as of the Interdisciplinary Center for Scientific Computing (IWR) at Heidelberg.

\bibliography{references}% Produces the bibliography via BibTeX.

\end{document}

%\preprint{APS/123-QED}

%\title{Thin plate elasticity of bent epithelial monolayers derived from a mean-field theory of a 3D vertex model}
\title{Supplemental material for:\\
Morphological instability at topological defects\\
in a three-dimensional vertex model for spherical epithelia}

\author{Oliver M. Drozdowski}
\affiliation{Institute for Theoretical Physics and BioQuant, Heidelberg University, 69120 Heidelberg, Germany}
\author{Ulrich S. Schwarz}%
 \email{Corresponding author: schwarz@thphys.uni-heidelberg.de}
\affiliation{Institute for Theoretical Physics and BioQuant, Heidelberg University, 69120 Heidelberg, Germany}

%\date{\today}% It is always \today, today,
             %  but any date may be explicitly specified

%\keywords{Suggested keywords}%Use showkeys class option if keyword
                              %display desired

\renewcommand{\theequation}{S\arabic{equation}}
\renewcommand{\thefigure}{S\arabic{figure}}      
\renewcommand{\thetable}{S\arabic{table}}   
                       
\maketitle
\onecolumngrid
%\tableofcontents
\section{Full derivation of the stretching energy density}
\begin{figure}[!ht]
	\centering
	\includegraphics[width=3.5in]{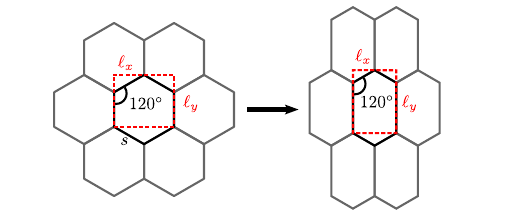}
	\caption{Parametrization of the mid-plane of a cell undergoing stretching. For the angular relaxed state the initial undeformed angles (left) do not change upon stretching (right) due to membrane tension.}
\label{fig:stretching_parametrization}
\end{figure}

We consider the dimensionless vertex model (VM) Hamiltonian with $\Gamma_l=1=V$:
\begin{equation}\label{eq:hamiltonian_vertex}
    E_\mathrm{VM}=\sum_\mathrm{cells} \left( \Gamma_\mathrm{a} A_\mathrm{a} + \Gamma_\mathrm{b} A_\mathrm{b} + \frac{1}{2} \sum_\text{lateral faces} A_\mathrm{l} \right)\ .
\end{equation}
Looking at the mid-plane, we can parametrize the area as shown in Fig.~\ref{fig:stretching_parametrization} and as proposed in Ref.~\cite{Staddon_SM23_Non_affine_deformations_VM}.
The cell has lengths $\ell_x$ and $\ell_y$ and an internal hexagonal angle of $2\pi/3$.

We do not consider bending in the derivation of the elastic constants and thus the apical and basal areas are identical to the mid-plane areas
\begin{equation}
A_\mathrm{a} = A_\mathrm{b} = \ell_x \ell_y.
\end{equation}
Due to membrane tension the internal angle is assumed to be unchanged when the mid-plane is stretched and as such we may parametrize the lateral areas via the total perimeter of the cell
\begin{equation}
P_\mathrm{tot} = 2\ell_y + \left( \frac{2+\cos(2\pi/3)}{\sin(2\pi/3)} \right) \ell_x.
\end{equation}
The lateral area is given by $A_\mathrm{l}=P_\mathrm{tot}h$.
For the reference state we may calculate the height and edge length
with minimal energy by minimizing Eq.~(\ref{eq:hamiltonian_vertex}) with respect to $h$, i.e.\
\begin{equation}\label{eq:equilibrium_cell_height_edge_length}
\begin{aligned}
    h_\mathrm{min} =& \frac{4^{2/3} \zeta_\mathrm{geom}^{1/3}}{n^{2/3}} (\Gamma_\mathrm{a}+\Gamma_\mathrm{b})^{2/3}, \\
    s_\mathrm{min} =& \frac{n^{1/3}}{4^{1/3} \zeta_\mathrm{geom}^{2/3}} ( \Gamma_\mathrm{a}+ \Gamma_\mathrm{b})^{-1/3},
\end{aligned}
\end{equation}
respectively. The geometric constant $\zeta_\mathrm{geom}$ links the edge length and the polygonal area (for the hexagon $\zeta_\mathrm{geom}=3^{3/2}/2$), i.e. $A_\text{a,b}=\zeta_\mathrm{geom}s^2$ with polygonal edge length $s$.

\begin{figure}[!ht]
	\centering
	\includegraphics[width=3.5in]{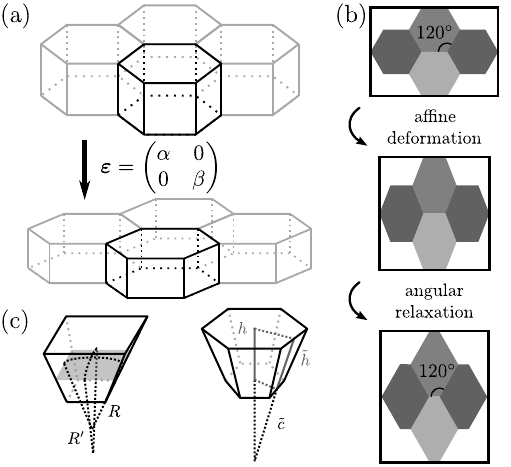}
	\caption{Deformations on a single cell level determine the stretching and bending energies. (a) We consider a strain tensor $\bm{\varepsilon}$ which varies slowly compared to the size of an individual cell. The deformation imposed on the cells changes the energy compared to the reference configuration. (b) Stretching with an affine lattice deformation is not consistent with force balance at mid-plane junctions connecting the membranes. Subsequent angular relaxation yields the minimal energy shape. (c) A cell is bent with two principal radii of curvature $R$ and $R'$, defining curvatures $c$ and $c'$, respectively. For non-rectangular cells the resulting energy is approximated via angular averaging, see text. The lateral face of the cell with height $h$ has trapezoidal height $\tilde{h}$ and experiences curvature $\tilde{c}$.}
\label{fig:deformation}
\end{figure}

The mid-plane is subject to a deformation described by the diagonalized strain tensor
\begin{equation}
    \bm{\varepsilon} =
    \begin{pmatrix}
        \alpha & 0 \\
        0 & \beta
    \end{pmatrix}.
\end{equation}
For the deformed state we may now look at deformed quantities $\ell_x', \ell_y', h'$, which are deformed via the principal strains $\alpha$ and $\beta$, cf.~Fig.~\ref{fig:deformation}(a), as such that
\begin{equation}
\begin{aligned}
\ell_x ' &= (1+\alpha)\ \ell_x, \\
\ell_y' &= (1+\beta)\ \ell_y, \\
h' &= \frac{h}{(1+\alpha)(1+\beta)}
\end{aligned}
\end{equation}
We have used the volume conservation $V=1$ here. The deformed energy for one cell can then be computed via
\begin{equation}\label{eq:full_stretching_energy}
E = (\Gamma_\mathrm{a} + \Gamma_\mathrm{b}) (1+\alpha) (1+\beta) \zeta_\mathrm{geom} s^2 + \frac{3}{2} sh \left( \frac{1}{1+\alpha} + \frac{1}{1+\beta} \right),
\end{equation}
with (undeformed) hexagonal edge length $s$.
Dividing by the mid-plane area $\zeta_\mathrm{geom}s^2$, inserting the reference height and edge length, Eq.~(\ref{eq:equilibrium_cell_height_edge_length}), 
and Taylor expanding with respect to $\alpha$ and $\beta$ to second order, we obtain the energy density
\begin{equation}
e = \frac{E}{\zeta_\mathrm{geom}s^2} = (\Gamma_\mathrm{a} + \Gamma_\mathrm{b}) \left( 3 + \alpha^2 + \beta^2 + \alpha \beta\right),
\end{equation}
which is the linear elasticity result considered in the main manuscript.

To compare to the energy without non-affine angular relaxation, cf.~Fig.~\ref{fig:deformation}(b), we consider the deformed lateral areas after affine transformation.
The apical and basal areas after deformation are identical to the case with angular relaxation: 
the deformed mid-plane area is $A_\mathrm{mid}'=|\det(1+\bm{\varepsilon})|A_\mathrm{mid}$. 
For the lateral faces we employ an angular averaging method: we first consider the edge vector $\mathbf{e}=s(\cos(\theta), \sin(\theta))$ 
with edge angle $\theta$.
Then the deformed lateral area is determined from the deformed edge length and volume conservation to read
\begin{equation}
    A_\mathrm{l}'(\theta) = \frac{\sqrt{(1+\alpha)^2\cos^2(\theta)+(1+\beta)^2\sin^2(\theta)}}{(1+\alpha)(1+\beta)} A_l.
\end{equation}
Thus, the total lateral area for one cell is given by
\begin{equation}
    A'_\mathrm{tot,l}(n)=\sum_{i=0}^{n-1} A_\mathrm{l}'\left(\frac{i\, \pi}{n}\right),
\end{equation}
and we find for the energy density
\begin{equation}
e_\mathrm{affine} = (\Gamma_\mathrm{a} + \Gamma_\mathrm{b}) \left(3 +  \frac{9}{8} \alpha^2 + \frac{9}{8} \beta^2 + \frac{3}{4} \alpha\beta\right).
\end{equation}
The behavior for $\alpha\neq\beta$ is different with and without angular relaxation. For the case $\alpha=\beta$, where isotropic stretching does not change the angles, we find that both energy densities match, as they should.

Note that our assumption on the angular orientation of our lattice with respect to the principal strain directions does not influence the resulting elastic constants.
The 6-fold symmetry in the stretching energies implies that the material can be described by exactly two elastic constants in two-dimensional linear elasticity and as such the stretching energy should be valid within the applicability range of linear elasticity for arbitrary directions.

\begin{figure}[!ht]
	\centering
	\includegraphics[width=3.5in]{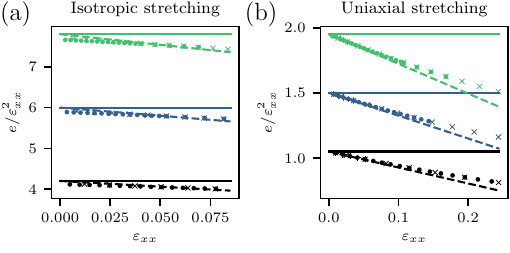}
	\caption{Comparison of the stretching energy densities of the numerical VM and the continuum model for isotropic stretching (a) and uniaxial stretching (b). Energy densities are devided by the quadratic strain to obtain the constant coefficient from linear elasticity. Symbols and colors are consistent to Fig. 3 in the main manuscript. Dashed and solid lines correspond to nonlinear cubic Taylor approximation and linear elasticity, respectively.}
\label{fig:stretching_energy}
\end{figure}

In the manuscript we have checked numerical results of the (effective) bulk modulus, Young's modulus and Poisson ratio based on the linear relationship between strain and stress.
For effects that depend both on bending and stretching we are usually interested in the stretching energy.
To show that our continuum model describes the numerical VM well, we have additionally compared the continuum and numerical stretching energy densities in Fig.~\ref{fig:stretching_energy}. 
We show the energy densities (per unit mid-plane area) divided by the quadratic strain as this is constant in linear elasticity with
\begin{equation}
e/\varepsilon_{xx}^2 = \frac{Y}{1-\nu} = 2B
\end{equation}
for isotropic stretching and
\begin{equation}
e/\varepsilon_{xx}^2 = Y/2
\end{equation}
for uniaxial stretching.

We do find similar results as in the moduli in the main manuscript. 
The nonlienar cubic model with coefficients as derived in the following section describes the energy well.
Linear elasticity describes the behavior for vanishing strain.
We find this behavior for both the isotropic and uniaxial stretching cases, cf.~Fig.~\ref{fig:stretching_energy}(a) and (b), respectively.

\section{Nonlinear treatment of the stretching energy}
In order to describe the VM in the framework of linear elasticity theory
we have considered the stretching
energy only up to quadratic order in the principal strains. To understand how the (effective) elastic
parameters change with increasing strain we now consider higher orders.
Going back to Eq.~(\ref{eq:full_stretching_energy}), we find for the full stretching energy density
\begin{equation}\label{eq:stretching_energy_density_full}
    e_\mathrm{stretch}^\mathrm{full} = (\Gamma_\mathrm{a} + \Gamma_\mathrm{b}) \left[ (1+\alpha)(1+\beta) + \frac{1}{1+\alpha} + \frac{1}{1+\beta} - 3\right].
\end{equation}

In nonlinear elasticity theory (NLET) we now consider the nonlinear strain tensor \cite{Howell_Ockendon_Solid_Mechanics}
\begin{equation}
    \mathcal{E}_{ij} = \frac{1}{2} (\partial_i u_j + \partial_j u_i + \partial_i u_k \partial_j u_k), 
\end{equation}
which means that the principal strains for a deformation in which the edges are stretched by
$1+\alpha$ and $1+\beta$, cf.~Fig.~\ref{fig:stretching_parametrization}, will be $\alpha+\alpha^2/2$ and $\beta+\beta^2/2$.
In NLET we consider the new coordinate of a material point $\chi(x,y)$ and must consider the deformation gradient tensor
$
    F_{ij} = \partial_{x_j}\chi_i
$.

The central quantity for the energy in NLET is the Cauchy-Green tensor
\begin{equation}
    \bm{\mathcal{C}} = \mathbf{F}^T\mathbf{F},
\end{equation}
which has eigenvalues $\lambda_x^2$ and $\lambda_y^2$ -- the squared principal stretches -- which are related to the principal strains as such that in our case
\begin{equation}
    \lambda_x = (1+\alpha),\quad \lambda_y = (1+\beta).
\end{equation}
We may now rewrite the stretching energy density, Eq.~(\ref{eq:stretching_energy_density_full}), in these principal stretches to find
\begin{equation}
    e_\mathrm{stretch}^\mathrm{full} = (\Gamma_\mathrm{a} + \Gamma_\mathrm{b}) \left[ \lambda_x\lambda_y + \frac{1}{\lambda_x} + \frac{1}{\lambda_y} - 3\right].
\end{equation}
Equivalently, we may look at the invariants of the Cauchy-Green tensor $I_1 = \mathrm{tr}(\bm{\mathcal{C}}) = \lambda_x^2 + \lambda_y^2$ and $I_2 = \det(\bm{\mathcal{C}})=\lambda_x^2\lambda_y^2$
to find
\begin{equation}
    e_\mathrm{stretch}^\mathrm{full} = (\Gamma_\mathrm{a} + \Gamma_\mathrm{b}) \left[ \sqrt{I_2} + \sqrt{\frac{I_1}{I_2} + \frac{2}{\sqrt{I_2}}} -3\right].
\end{equation}
This energy density does not match the standard set of widely used hyperelastic materials such as Neo-Hookean or Ogden.

To simplify our discussion of the nonlinearities of the elastic coefficients 
we now consider the next-higher order in the Taylor approximation of the energy density, Eq.~(\ref{eq:full_stretching_energy}),
\begin{equation}
    e_\mathrm{stretch}^\mathrm{cubic} = (\Gamma_\mathrm{a} + \Gamma_\mathrm{b}) \left[ \alpha\beta + \alpha^2 + \beta^2 - \alpha^3 - \beta^3 \right].
\end{equation}

For isotropic stretching we have $\alpha=\beta=\varepsilon_{xx}$ and thus for the energy density in 
both linear elasticity and our cubic Taylor approximation
\begin{equation}
\begin{aligned}
    e =& 2 (\mu+\lambda) \varepsilon_{xx}^2 = 2B \varepsilon_{xx}^2 \\
    \equiv& (\Gamma_\mathrm{a} + \Gamma_\mathrm{b}) (3-2\varepsilon_{xx}) \varepsilon_{xx}^2,
\end{aligned}
\end{equation}
i.e.\
\begin{equation}
B = \left(\frac{3}{2}-\varepsilon_{xx}\right) (\Gamma_\mathrm{a}+\Gamma_\mathrm{b}),  
\end{equation}
which is shown in the main manuscript.

For uniaxial stretching we have $\alpha=\varepsilon_{xx}$ and $\beta=-\nu \varepsilon_{xx}$.
Inserting this again into the energy densities of our Taylor approximation, and of linear elasticity yields
\begin{equation}
    \begin{aligned}
        e =& \frac{Y}{2} \varepsilon_{xx}^2 \\
        \equiv& (\Gamma_\mathrm{a} + \Gamma_\mathrm{b}) \left[ (1-\nu + \nu^2 + \nu^3 \varepsilon_{xx} - \varepsilon_{xx}) \varepsilon_{xx}^2\right].
    \end{aligned}
\end{equation}
The orthogonal strain $\varepsilon_{yy}=-\nu\varepsilon_{xx}$ and thus also the Poisson ratio minimize the stretching energy (density) for a given $\varepsilon_{xx}$.
We may differentiate the energy density with respect to $\nu$ to find this minimum and get
\begin{equation}
    -1 + 2 \nu + 3 \nu^2 \varepsilon_{xx} = 0.
\end{equation}
We want to find the linear dependence of $\nu$ on the strain, i.e.\ we use the ansatz $\nu = 1/2 + \tilde{\nu} \varepsilon_{xx}$, and find
\begin{equation}
    2 \tilde{\nu} \varepsilon_{xx} + \frac{3}{4} \varepsilon_{xx} + \mathcal{O}(\varepsilon_{xx}^2) = 0,
\end{equation}
where we neglected orders quadratic and higher in the strain, i.e.\ 
\begin{equation}
    \nu=\frac{1}{2} - \frac{3}{8} \varepsilon_{xx}.
\end{equation}
Inserting this into the energy density we also find 
\begin{equation}
    Y=\left(\frac{3}2 - \frac{7}{4}\varepsilon_{xx}\right) (\Gamma_\mathrm{a}+\Gamma_\mathrm{b}),
\end{equation}
which are the functions depicted in the main manuscript.

\section{Full derivation of mean-field bending constants}
We consider the non-dimensional Hamiltonian for a single cell, Eq.~(\ref{eq:hamiltonian_vertex}),
and assume principal curvatures $c$ and $c'$ with corresponding radii $R$ and $R'$ respectively.
The center height $h$ is assumed to not change when the cell is bent, similiarly to Fig.~\ref{fig:deformation}(c),
Assuming that a mid-plane edge is parallel to the axis with curvature $c$, 
the lateral face will form a trapezoid and we find
for the apical and basal edge lengths, $a$ and $b$, respectively,
\begin{equation}
\begin{aligned}
a &= (1+ch)\ s, \\
b &= (1-ch)\ s.
\end{aligned}
\end{equation}
Note that $s=(a+b)/2$. 
We have assumed the basal edge length to be smaller for positive curvature here.
As the apical and basal areas are related to edge lengths in both perpendicular directions we may write for the areas
\begin{equation}
\begin{aligned}
A_\mathrm{a} &= \left(1+\frac{ch}{2}\right)\left(1+\frac{c'h}{2}\right)\ \zeta_\mathrm{geom}s, \\
A_\mathrm{b} &= \left(1-\frac{ch}{2}\right)\left(1-\frac{c'h}{2}\right)\ \zeta_\mathrm{geom}s.
\end{aligned}
\end{equation}
Noticing that for a cut through the cell at a height of $z$ with respect to the mid-plane the areas can be written accordingly, we find for the bent cell volume
\begin{equation}\label{eq:volume_conservation}
V=\left(1+\frac{cc'h^2}{12}\right) h A_\mathrm{mid},
\end{equation}
with mid-plane area $A_\mathrm{mid}=\zeta_\mathrm{geom}s$.
For the derivation of the lateral area we consider the curvature orthogonal to the edge $\tilde{c}$ and introduce the geometrical constant $\eta_\mathrm{geom}$, which relates the polygonal edge length to the in-radius (for hexagons $\eta_\mathrm{geom}=\sqrt{3}/2$). , which we will denote as $t_s$ for the mid-plane.
On the apical and basal side we have
\begin{equation}
t_a = \left(1 + \frac{\tilde{c}h}{2}\right) t_s, \qquad t_b = \left(1- \frac{\tilde{c}h}{2}\right)t_s,
\end{equation}
respectively.
The trapezoidal height of the lateral face is
\begin{equation}\label{eq:trapezoidal_height}
\tilde{h}(\tilde{c}) = \sqrt{h^2 + (t_a - t_b)^2} = \sqrt{h^2 + \tilde{c}^2 h^2\ t_s^2}
\end{equation}
To determine the total lateral area we employ an angular averaging scheme:
we consider the curvature in the direction with angle $\theta$ from the direction of principal curvature $c$
\begin{equation}
\tilde{c}(\theta) = \cos^2(\theta) c + \sin^2(\theta) c',
\end{equation}
and average the curvature over the $n$-fold symmetry.
The total deformed lateral area is thus
\begin{equation}
    A'_\mathrm{tot,l}(n) = \sum_{i=0}^{n-1} s \tilde{h}\left(\tilde{c}\left(\frac{i\, \pi}{n}\right)\right),
\end{equation}
where the deformed mid-plane edge length $s$ is determined via volume conservation, 
Eq.~(\ref{eq:volume_conservation}), 
and the trapezoidal height is given by Eq.~(\ref{eq:trapezoidal_height}).

The apical and basal energy for a bent cell is
\begin{equation}
E_\mathrm{ab} = \left( \Gamma_\mathrm{a} + \Gamma_\mathrm{b}\right) \left( 1+ \frac{cc'h^2}{4} \right) \left( 1+ \frac{cc'h^2}{12}\right)^{-1} A_\mathrm{mid}^{(0)} +
\left(\Gamma_\mathrm{a} - \Gamma_\mathrm{b}\right) \frac{c+c'}{2} h A_\mathrm{mid}^{(0)}.
\end{equation}

The lateral bent energy for the hexagon is
\begin{equation}
E_\mathrm{lat} = \frac{1}{2} \left(1+\frac{cc'h^2}{12}\right)^{-1/2} h^{-1/2} \zeta_\mathrm{geom}^{-1/2}
\sum_{i=0}^{5} \sqrt{h^2 + \frac{\tilde{c}^2(i\pi/3)\ h^2}{1+cc'h^2/12}\ \frac{\eta_\mathrm{geom}^2}{\zeta_\mathrm{geom}\ h} }.
\end{equation}

Note that these equations are exact for $c=c'$ and for rectangular cells, but not necessarily for other polygonal shapes.
We now divide by the undeformed mid-plane area, $A_\mathrm{mid}^{(0)}=1/h$, to obtain the energy density and perform a Taylor expansion for small $ch$ and $c'h$ up to second order. We find

\begin{equation}
\begin{aligned}
e =& (\Gamma_\mathrm{a} + \Gamma_\mathrm{b})\ \left( 1+ \frac{cc'h^2}{6} \right) +(\Gamma_\mathrm{a} - \Gamma_\mathrm{b}) \frac{c+c'}{2} h \\
&+ \frac{3 h^{3/2}}{\zeta_\mathrm{geom}^{1/2}} + \frac{\eta_\mathrm{geom}^2 h^{1/2}}{\zeta_\mathrm{geom}^{3/2}} \left( \frac{9}{16} c^2 + \frac{9}{16} c'^2 +  \frac{6}{16} cc'\right)
- \frac{h^{7/2}}{\zeta_\mathrm{geom}^{1/2}}\ \frac{cc'}{8} \\
=& (\Gamma_\mathrm{a} + \Gamma_\mathrm{b}) + \frac{3 h^{3/2}}{\zeta_\mathrm{geom}^{1/2}} - (\Gamma_\mathrm{b}-\Gamma_\mathrm{a})^2\ \frac{h^{3/2}\zeta_\mathrm{geom}^{3/2}}{9 \eta_\mathrm{geom}^2} \\
&+ \frac{9}{16}\ \frac{\eta_\mathrm{geom}^2 h^{1/2}}{\zeta_\mathrm{geom}^{3/2}} \left[ (c+c') - \frac{4}{9}\ \frac{h^{1/2} \zeta_\mathrm{geom}^{3/2}}{\eta_\mathrm{geom}^2} (\Gamma_\mathrm{b}-\Gamma_\mathrm{a}) \right]^2 \\
&+ \left[ \frac{(\Gamma_\mathrm{a}+\Gamma_\mathrm{b}) h^2}{6} - \frac{3}{4}\ \frac{\eta_\mathrm{geom}^2 h^{1/2}}{\zeta_\mathrm{geom}^{3/2}} - \frac{h^{7/2}}{8 \zeta_\mathrm{geom}^{1/2}} \right] cc',
\end{aligned}
\end{equation}
where we rewrote the result to compare it to the bending energy, Eq.~(4) in the main manuscript. The first line then corresponds to a mid-plane surface tension energy of the reference state, 
the second to the mean-curvature contribution to the bending energy, and the last line to the Gauss curvature contribution.
We identify
\begin{equation}
\begin{aligned}
\kappa =& \frac{9}{8}\  \frac{\eta_\mathrm{geom}^2 h^{1/2}}{\zeta_\mathrm{geom}^{3/2}}, \\
\kappa_\mathrm{G} =&  \frac{(\Gamma_\mathrm{a}+\Gamma_\mathrm{b}) h^2}{6} - \frac{3}{4}\ \frac{\eta_\mathrm{geom}^2 h^{1/2}}{\zeta_\mathrm{geom}^{3/2}} - \frac{h^{7/2}}{8 \zeta_\mathrm{geom}^{1/2}}, \\
c_0 =& \frac{4}{9}\ \frac{h^{1/2} \zeta_\mathrm{geom}^{3/2}}{\eta_\mathrm{geom}^2} (\Gamma_\mathrm{b}-\Gamma_\mathrm{a}),
\end{aligned}
\end{equation}
for the bending rigidity, saddle splay modulus and spontaneous curvature, respectively.

\begin{figure}[!ht]
	\centering
	\includegraphics[width=3.5in]{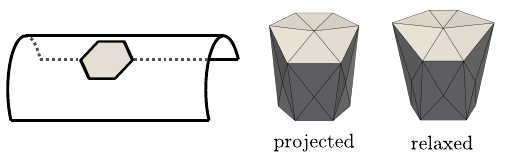}
	\caption{Non-affine transformations for cylindrical bending. Projecting a cell onto a cylindrical deformation (left) does not give the equilibrium cell shape. Further relaxation leads to non-affine apical/basal deformations (right).}
\label{fig:bending_nonaffine}
\end{figure}

Similarly as for uniaxial stretching, we have found that additional corrections of the coefficients are necessary due to non-affine deformations. 
Non-affine apical and basal deformations indeed justify such a correction in the case $c\neq c'$. 
In Fig.~\ref{fig:bending_nonaffine} we show cylindrically bent cells in the VM with nodes projected onto the surface before and after energy minimization.
These apical and basal deformations are similar to the deformations in uniaxial mid-plane stretching, cf.~Fig.~\ref{fig:deformation}(b).
We thus introduce correction factors $k_c, k_\mathrm{G}$ and $k_0$ for the bending rigidity, saddle splay modulus and spontaneous curvature, respectively, such that
\begin{equation}
\kappa \rightarrow k_c \kappa, \qquad \kappa_\mathrm{G} \rightarrow k_\mathrm{G} \kappa_G, \qquad c_0 \rightarrow k_0 c_0.
\end{equation}

Such non-affine deformations will only be present for cases with $c\neq c'$ and should not occur for $c=c'$, so the total energy density should be the same with and without corrections for a spherical deformation with $c=c'$.

We fix $k_c$ based on numerical results for bent plates with one vanishing curvature.
From the equal energies in the case of $c=c'$ we can derive for the saddle splay modulus correction
\begin{equation}
    k_\mathrm{G} = 1+ 2(1-k_c) \frac{\kappa}{\kappa_\mathrm{G}},
\end{equation}
i.e.\
\begin{equation}
\begin{aligned}
\kappa =& \frac{9 k_c}{8}\  \frac{\eta_\mathrm{geom}^2 h^{1/2}}{\zeta_\mathrm{geom}^{3/2}}, \\
\kappa_\mathrm{G} =& \frac{(\Gamma_\mathrm{a}+\Gamma_\mathrm{b}) h^2}{6} + \left(\frac{3}{2}- \frac{9}{4}k_c\right) \frac{\eta_\mathrm{geom}^2 h^{1/2}}{\zeta_\mathrm{geom}^{3/2}} - \frac{h^{7/2}}{8 \zeta_\mathrm{geom}^{1/2}}.
\end{aligned}
\end{equation}
To obtain the results in the main manuscript, we now insert the reference height (ignoring height changes due to stretching) $h_\mathrm{min}$, Eq.~(\ref{eq:equilibrium_cell_height_edge_length}), and the geometric constants for hexagons.

For the spontaneous curvature computing such a correction is not as simply done.
Assuming the bending energies with and without corrections are identical for $c=c'$, leads to a quadratic equation for the $k_0$, i.e.\
\begin{equation}
    k_0^2 - \left( 4\frac{c}{c_0}\right) k_0 + \left( 4 \frac{c}{c_0} - 1\right) \frac{1}{k_c}=0,
\end{equation}
which has the solution of $k_0=1$ for $k_c=1$ as it should.
With the correction in the mean curvature, $k_c\neq 0$, we obtain
\begin{equation}
    k_{0\pm} = 2 \frac{c}{c_0} \pm \sqrt{4 \frac{c^2}{c_0^2} + \left(1-4\frac{c}{c_0}\right) \frac{1}{k_c}}.
\end{equation}
For $k_c=1$ we see that $k_{0+}$ is valid for $c_0<2c$ and $k_{0-}$ for $c_0>2c$, but for $k_c\neq 1$ the continuity of $k_0$ is lost if the curvature matches the spontaneous curvature here.

As such it makes sense to assume that the curvature at which the minimum in the mean-curvature bending energy occurs is not shifted by the correction, which implies $k_0=1$. 
However, this will lead to energetic changes.
These are a consequence of the assumption of small $ch,ch'$ in the above derivation,
which will not properly work for $c_0\neq0$.
Future work with $\Gamma_\mathrm{a}\neq \Gamma_\mathrm{b}$ could investigate possible expansions at $H\approx c_0$ insted of $H\approx 0$ to mitigate this issue.
This is, however, beyond the scope of this work.

\section{Numerical vertex model implementation}

\subsection*{Formulation of the vertex model}

\begin{figure}[!ht]
	\centering
	\includegraphics[width=3.5in]{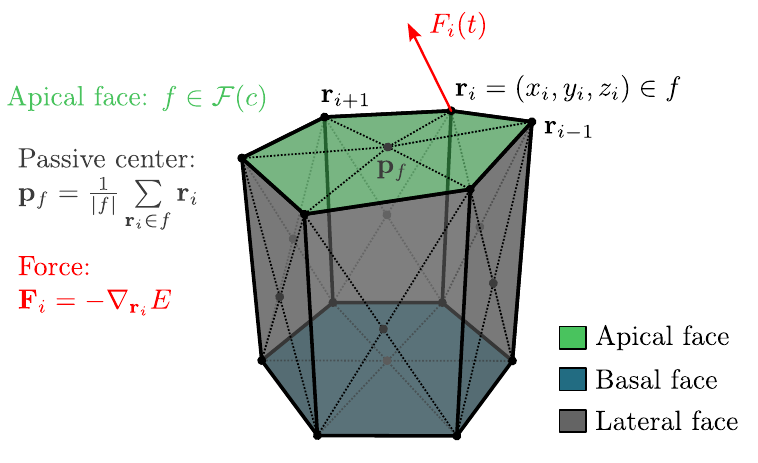}
	\caption{Schematic representation of a cell in the vertex model. The cell is described as a polyhedron with boundary faces. The boundary faces are described by nodes and are triangulated using a passive center node. Forces that are used for energy minimization are calculated from the gradient of the total energy.}
\label{fig:vertex_model}
\end{figure}

The three-dimensional vertex model (VM) is a collection of polyhedral cells $\mathcal{C}$ with corresponding faces $\mathcal{F}(c)$ for $c\in\mathcal{C}$.
The faces are a collection of nodes $\mathbf{r}_i\in\mathbb{R}^3$, see Fig.~\ref{fig:vertex_model}, and can be separated into three different classes: apical, basal and lateral faces.
We assume the apical and basal network topologies to be identical, 
i.e.\ the lateral faces are always (possibly non-coplanar) rectangles 
with four nodes and two lateral edges connecting the corresponding apical and basal nodes.
To calculate areas and volumes consistently in the case of non-coplanar face nodes, we triangulate 
each face with a passive center node, as suggested in Ref.~\cite{Krajnc_PRE18_VM_fluidization},
\begin{equation}
    \mathbf{p}_f = \frac{1}{|f|} \sum_{\mathbf{r}_i\in f} \mathbf{r}_i,
\end{equation}
where $|f|$ denotes the number of nodes in face $f$.

The energy of the VM is given by Eq.~(\ref{eq:hamiltonian_vertex}) -- in the numerical calculation we also set $V=1=\Gamma_\mathrm{l}$.
As we want to minimize the total energy we compute a force $\mathbf{F}_i$ for node $\mathbf{r}_i$, which is given as the negative gradient of the energy with respect to the node position,
\begin{equation}\label{eq:overdamped_eom_nodes}
    \mathbf{F}_i = - \nabla_{\mathbf{r}_i} E.
\end{equation}
Note that non-neighboring nodes can contribute via the passive center, which has to be accounted for via the chain rule.
To minimize the energy we assume the nodes' dynamics to follow overdamped dynamics, i.e.\
\begin{equation}
    \frac{\mathrm{d}}{\mathrm{d}t} \mathbf{r}_i = \mathbf{F}_i,
\end{equation}
where we assumed a mobility (or viscosity) of one.

We implemented the VM in the cell-based package of Chaste \cite{Chaste_Software_2020} by adding the three-dimensional geometry and the corresponding Hamiltonian.
The equation of motion for the nodes, Eq.~(\ref{eq:overdamped_eom_nodes}), is integrated using Euler stepping.
To achieve volume conservation we project the forces (including edge forces) on the subspace of volume-conserving forces and correct for possible volume changes \cite{Krajnc_PRE18_VM_fluidization, Brakke_92_Surface_Evolver}.

\subsection*{Edge forces and boundary conditions}
When considering a plate of size $(n_x, n_y)$, we have edge lengths of
\begin{equation}
    L_x = \sqrt{3}\ n_x s,\qquad L_y = \frac{3}{2}\ n_y s,
\end{equation}
with polygonal edge length $s$.
Edge stresses are then considered as an additional force. 
The total force on the edge is calculated by multiplying the stress with either $L_x$ or $L_y$. 
This total force is then equally distributed to all edge nodes, which have a corresponding opposing edge node in the hexagonal lattice.
To maintain a stable compressive state, the forces are implemented as normal forces to the edge.
The force directions at opposing edges are then parellelized to not allow for slipping, i.e.\ not have a net-force resulting in rigid body movement.

At the lateral edge faces the surface tension formula is not modified, because the tension is not shared by two neighboring cells. 
These faces effectively have half the regular lateral surface tensions.
Lateral edges, i.e.\ the edges connecting oppsoing apical and basal nodes, at the plate boundary are constrained to be parallel to the averaged normal vector of the apical and basal faces of the cells to which the edge belongs.
This allows for an orthogonal lateral edge at the boundaries.

We used orthogonal boundaries for finite-size plates, 
i.e.\ lateral edges are fixed to be parallel to the averaged apico-basal normal vectors. 
Simply-supported boundary conditions are implemented by constraining the lateral edges' centers of the outer most nodes on the edges to the initial mid-plane surface ($z=h/2$).
If the edges do not fulfill this condition, the corresponding nodes are moved accordingly at every integration step.
This allows for a non-vanishing derivative of the deflection function at the boundaries but constrains the mid-plane position.
To implement cylindrically or spherically bent plates 
all lateral edge centers are constrained to the surface of the corresponding deflection function.
To implement straight boundaries of the rectangular plate the lateral edge centers are enforced to be in the corresponding plane defined by the boundary normal 
and the center of mass of all the outer-most lateral edges.

\subsection*{Icosahedral sphere generation}
To generate spheres with icosahedral symmetry the center nodes of a corresponding icosahedron, following the Caspar-Klug construction, are projected onto a sphere. 
We then create the Delaunay triangulation on the sphere by taking the convex hull \cite{CGAL}, which yields the face centers as triangulation points.
The corresponding Voronoi vertices are then computed as the circumcenters of the Delaunay triangles on the sphere and serve as the nodes on the luminal side of the sphere.
The outer side is created by duplicating the nodes, increasing the radius, and constructing the final mesh.
All icosahedral spheres are initialized on the spherical surface and then the energy is minimized.

\subsection*{Spontaneous curvature patterning}
For the spontaneous curvature patterning the apical and the basal surface tensions of the pentagonal defect 
cells and their hexagonal nearest neighbors were chosen as such that the sum of the tensions is unchanged, i.e.
\begin{equation}
    \Gamma_\text{a}=\Gamma + \frac{\Delta\Gamma}{2},\qquad \Gamma_\text{b}=\Gamma - \frac{\Delta\Gamma}{2},
\end{equation}
where we used $\Gamma=1.1$ in the manuscript. This will induce a spontaneous curvature without changing the bending rigidities or the elastic moduli in the cells. The apical side is assumed to be luminal, i.e. toward the inside of the (icosahedral) sphere, matching the polarity in experimental situations such as intestinal organoids growing in three-dimensional culture.

\section{Finite element method for spherically bent rectangular plate}
Since the deflection function for a spherical deformation with radius $R$ is known, i.e.\
\begin{equation}
    f(x,y) = R - \sqrt{R^2 - x^2 -y^2},
\end{equation}
we can solve the thin plate theory, with strain 
\begin{equation}\label{eq:strain_moederate_bending_deflection}
    \varepsilon_{ij} = \frac{1}{2} \left( \partial_i u_j + \partial_j u_i + \partial_i f \partial_j f \right)
\end{equation}
as we would solve the in-plane problem of linear elasticity.

We want to solve the following system of equations on $\Omega = [-L_x/2, L_x/2]\times[-L_y/2,L_y/2]$
\begin{equation}
    \begin{aligned}
        -\partial_i \sigma_{ij} = & 0, \\
        \sigma_{ij}(\bm{\varepsilon}) = & 2 \mu \varepsilon_{ij} + \lambda \varepsilon_{kk} \delta_{ij}, \\
        \varepsilon_{ij}(\mathbf{u},f)  = & \frac{1}{2} \left( \partial_i u_j + \partial_j u_i + \partial_i f \partial_j f \right) = \tilde{\varepsilon}_{ij}(\mathbf{u}) + \tilde{f}_{ij}.
    \end{aligned}
\end{equation}
with stress-free edges and $u_x(0,0)=u_y(0,0)=0$. We have split the strain tensor into the component resulting from the in-plane deformation $\tilde{\varepsilon}_{ij}(\mathbf{u})=(\partial_i u_j + \partial_j u_i)/2$ and the component from the deflection function $\tilde{f}_{ij}=(\partial_i f \partial_j f)/2$ for notational convenience.

To obtain a weak formulation for a solution with the finite element method we introduce a vector valued test function $\mathbf{v}$ and consider
\begin{equation}
    -\int_\Omega \partial_i \sigma_{ij} v_j\, \mathrm{d}S = \int_\Omega \sigma_{ij} (\partial_i v_j)\, \mathrm{d}S,
\end{equation}
where we used partial integration and the boundary condition.

This can now be rewritten into the variational formulation, i.e.\ find $\mathbf{u}$ in a suitable vector-valued finite element space such that for all $\mathbf{v}$
\begin{equation}
    a(\mathbf{u},\mathbf{v}) = L(\mathbf{v}).
\end{equation}
The bilinear form for our problem reads
\begin{equation}
    a(\mathbf{u},\mathbf{v}) = \int_\Omega \sigma_{ij}(\tilde{\bm{\varepsilon}}(\mathbf{u}))\, \tilde{\varepsilon}_{ij}(\mathbf{v})\, \mathrm{d}S
\end{equation}
The right hand side contains all the other terms from the deflection function
\begin{equation}
    L(\mathbf{v}) = - \int_\Omega \sigma_{ij}(\tilde{\bm{f}})\, \tilde{\varepsilon}_{ij}(\mathbf{v})\, \mathrm{d}S.
\end{equation}

This linear system was solved in FEniCS \cite{Alnaes_2015_FEniCS} for given $\Omega$ and $R$ where for both $\mathbf{u}$ and $\mathbf{v}$ linear Lagrange elements were chosen. 

\subsection*{Comparison of continuum and VM height distributions in bent monolayers}

\begin{figure}[!ht]
	\centering
	\includegraphics[width=3.5in]{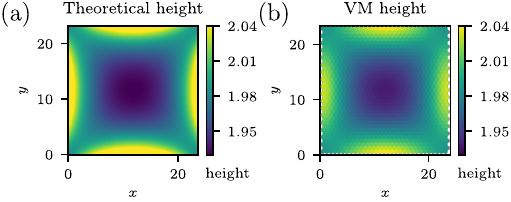}
	\caption{Bending of a rectangular 3D VM monolayer. (a) Cell height for a spherically bent rectangular monolayer with $\Gamma=1.3$, size $(31,35)$ and $c=1/20$ from finite element energy minimization of the continuum theory. (b) The same
    for the 3D VM in the undeformed mid-plane.}
\label{fig:si_bending_height}
\end{figure}

Due to the volume conservation mid-plane area changes are related to height and we may calculate the deformed height $h'$ via $h'=h/\det(1+\bm{\varepsilon})$.
Fig.~\ref{fig:si_bending_height}(a) shows the expected height, 
while the bent 3D VM monolayer cell heights are shown in (b).
The continuum model predicts the height distribution well,
but for the strongly bent monolayer shown in Fig.~\ref{fig:si_bending_height} it slightly underestimates the compression at the edge centers.

\section{Post-buckling approximation for compressed rectangular simply-supported plate}
For a simply-supported rectangular plate, compressed along the edges parallel to the $y$-direction with compressive stress $\sigma_{xx}$,
the critical buckling stress is given by \cite{Timoshenko1961}
\begin{equation}\label{eq:critical_buckling_stress}
\sigma_\mathrm{crit} = \frac{\pi^2 \kappa}{L_x^2} \left( m + \frac{1}{m}\, \frac{L_x^2}{L_y^2}\right)^2
\end{equation}
where the multiplicity $m$ describes the number of half-waves along the $x$-direction and it is chosen such that the critical stress is minimal, e.g.\ $m=1$ for quadratic plates. 

To approximate the post-buckling behavior, we consider a rectangular plate with straight edges (i.e. it is in a movable but straight frame).
We introduce the half-lengths $\tilde{L}_x = L_x/2$ and $\tilde{L}_y = L_x/2$ and choose the center of the plate as the zero in the coordinate system.
The rectangular plate is compressed along the $x$-direction, i.e.\ it is loaded along the edges $x=\pm\tilde{L}_x$.

Solving the post-buckling behavior exactly is very difficult and can only be achieved numerically \cite{Rhodes_03_Post_buckling_thin_plates}.
However, an approximation is possible based on the dominant mode in the Fourier series description of the deformations.
The in-plane deformations $\mathbf{u}$ and the deflection function can be assumed to be \cite{Timoshenko1961}
\begin{equation}
\begin{aligned}
    u_x &= C_x \sin\left(\frac{\pi x}{\tilde{L}_x}\right) \cos\left(\frac{\pi y}{2\tilde{L}_y}\right) - e x \\
    u_y &= C_y \sin\left(\frac{\pi y}{\tilde{L}_y}\right) \cos\left(\frac{\pi x}{2\tilde{L}_x}\right) + a y \\
    f &= A \cos\left( \frac{\pi x}{2\tilde{L}_x} \right) \cos\left( \frac{\pi y}{2\tilde{L}_y} \right).
\end{aligned}
\end{equation}
where $e$ is given by the strain in $x$-direction from the compression, i.e.\ $e=-\varepsilon_{xx}=\sigma_{xx}/Y$ with compressive stress $\sigma_{xx}$.
The constant $a$ is determined from the condition that the integrated stress along the unloaded edge vanishes and is given by
\begin{equation}
    a = -\frac{\pi^2 A^2}{16 \tilde{L}_y^2} + \frac{2C_y}{\tilde{L}_y} + \nu e.
\end{equation}
The deflection amplitude $A$ and the in-plane deformation amplitudes $C_x$ and $C_y$ are unknowns and will be determined via energy minimization.

The stresses can now be determined in the framework of moderately bent plates, cf.~Eq.~(\ref{eq:strain_moederate_bending_deflection}).
The stretching energy can be computed via the energy density integral
\begin{equation}
    E_\mathrm{stretch} = \int_{-\tilde{L}_x}^{\tilde{L}_x}\mathrm{d}x \int_{-\tilde{L}_y}^{\tilde{L}_y}\mathrm{d}y\ \frac{1}{2} \left( \frac{Y}{1+\nu} \varepsilon_{ij}^2 + \frac{Y\nu}{1-\nu^2} \varepsilon_{kk}^2 \right).
\end{equation}
For the bending energy one can show that the Gauss curvature contribution vanishes for the given deflection function and one finds \cite{Timoshenko1961}
\begin{equation}
    E_\mathrm{bending} = \frac{\pi^4 \tilde{L}_x \tilde{L}_y A^2 \kappa}{32} \left( \frac{1}{\tilde{L}_x^2} + \frac{1}{\tilde{L}_x^2} \right)^2.
\end{equation}
The total energy is thus
\begin{equation}\label{eq:full_energy_bifurcation}
\begin{aligned}
    E =\ & E_\mathrm{stretch} + E_\mathrm{bending} \\ 
    =\ & 2 e^2 \tilde{L}_x \tilde{L}_y Y + \frac{8 C_x C_y (1+\nu) Y}{9(1-\nu^2)} + \frac{C_x^2 \tilde{L}_y \pi^2 Y}{2 \tilde{L}_x (1-\nu^2)} + \frac{C_y^2 \tilde{L}_y (1-\nu) \pi^2 Y}{16 \tilde{L}_x (1-\nu^2)} \\
    & \qquad\quad + \frac{\tilde{L}_x \left[C_x^2(1-\nu)\pi^2 + 8 C_y^2 (16+\pi^2)\right] Y}{16\tilde{L}_y (1-\nu^2)} \\
    &+ A^2 \left\{ \frac{1}{32} \kappa \tilde{L}_x \tilde{L}_y \pi^4 \left(\frac{1}{\tilde{L}_x^2} + \frac{1}{\tilde{L}_y^2}\right) - \frac{5 C_y \tilde{L}_x \pi^2 Y}{12 \tilde{L}_y^2 (1-\nu^2)} - \frac{C_x \tilde{L}_y \pi^2 Y}{6 \tilde{L}_x^2 (1-\nu^2)} \right. \\
    & \qquad\quad \left. - \frac{e \tilde{L}_y \pi^2 Y}{8 \tilde{L}_x} - \frac{C_x (1-3\nu) \pi^2 Y}{24 \tilde{L}_y (1-\nu^2)} -\frac{C_y (1-9\nu) \pi^2 Y}{24 \tilde{L}_x (1-\nu^2)} \right\} \\
    &+ A^4 \left\{ \frac{9 \tilde{L}_x \pi^4 Y}{2048 \tilde{L}_y^3 (1-\nu^2)} + \frac{9 \tilde{L}_y \pi^4 Y}{2048 \tilde{L}_x^3 (1-\nu^2)} + \frac{(1-8\nu) \pi^4 Y}{1024 \tilde{L}_x \tilde{L}_y (1-\nu^2)} \right\}.
\end{aligned}
\end{equation}

This is a fourth-order polynomial equation in $A$. 
The minimum of this energy function is then determined by the zeros of the derivative with respect to the different parameters.
Due to the form with respect to $A$ we then obtain the normal form of a pitchfork bifurcation.
As this function is quite involved, we determine the minimum numerically by minimizing the energy with respect to the parameters $A$, $C_x$ and $C_y$.

In order to compare the VM results with this post-buckling bifurcation we need to additionally consider that the in-plane deformations $\mathbf{u}$ are considered inside the $z=0$ plane due to the assumption of moderate plate bending and not inside the real buckled mid-plane.
For thin elastic plates this yields a discontinuous transition to a smaller (in-xy-plane) stiffness in the post-buckling regime, which is known to be approximately \cite{Rhodes_03_Post_buckling_thin_plates}
\begin{equation}
    Y_\mathrm{post} \approx Y/2.
\end{equation}

\begin{figure}[!ht]
	\centering
	\includegraphics[width=3.5in]{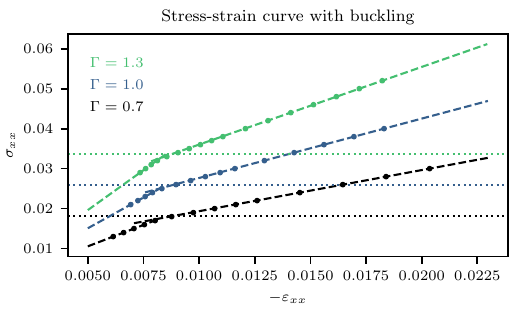}
	\caption{Stress-strain curves for the buckling rectangular VM monolayer from Fig.~4(a) in the main manuscript. The pre- and post-buckling behavior is linear with different slopes (Young's moduli). The dashed lines are linear fitting results, summarized in Table~\ref{tab:youngs_correction} and the horizontal dotted lines are the theoretical buckling thresholds.}
\label{fig:buckling_stiffness}
\end{figure}

\newcolumntype{C}[1]{>{\hsize=#1\hsize\centering\arraybackslash}X}%
\begin{table}[!ht]
\begin{tabularx}{0.55\textwidth}{ C{0.1} C{0.3} C{0.3} C{0.2} }
\hline \hline
$\Gamma$ & Pre-buckling slope & Post-buckling slope & $k_Y$\\
\hline
$0.7$ & $2.144$ & $1.029$ & $0.480$\\
$1.0$ & $3.073$ & $1.485$ & $0.483$\\
$1.3$ & $3.994$ & $1.942$ & $0.486$\\
\hline \hline
\end{tabularx}
\caption{\label{tab:youngs_correction}Fitting-results for pre-buckling and post-buckling slopes and post-buckling correction factor for Young's modulus $k_Y$.}
\end{table}

To verify this post-buckling behavior in our VM simulations, we show the compressive stress as function of $e=-\varepsilon_{xx}$ for the data from Fig.~4 in the main manuscript in Fig.~\ref{fig:buckling_stiffness}.
We observe that indeed the Young's modulus, i.e. the slope, of the VM can be described by two linear functions.
We used linear regression on the first three points and the last four points for each $\Gamma$ in Fig.~\ref{fig:buckling_stiffness} to determine the slopes before and after buckling 
and calculated the numerical post-buckling stiffness correction $k_Y$ as the ratio of the buckled and non-buckled slopes, cf.~Table~\ref{tab:youngs_correction}.
Using $Y_\mathrm{post} = k_Y Y$ we then determined the bifurcation diagram depicted in Fig.~4 in the main manuscript by calculating $e$ from the applied stress and 
minimizing the energy, Eq.~(\ref{eq:full_energy_bifurcation}), to determine $A(\sigma_{xx})$.

To show the agreement of the proposed deformation fields with the main characteristic features of the compressed VM plate,
we can consider the deformed height of the cells.
It is related to the (hydrodynamic) area-changing strain via $h'=h/\det(1+\bm{\varepsilon})$.
We find
\begin{equation}\label{eq:height_det_buckled}
\begin{aligned}
    \det(1+\bm{\varepsilon}) =& \left[1+ \frac{2C_y}{\tilde{L}_y} + e\nu - \frac{A^2 \pi^2}{16 \tilde{L}_y^2} + \frac{C_y \pi \cos(\pi y/\tilde{L}_y) \cos(\pi x/2\tilde{L}_x)}{\tilde{L}_y} + \frac{A^2\pi^2 \cos^2(\pi x/ 2\tilde{L}_x) \sin^2(\pi y/2\tilde{L}_y)}{8 \tilde{L}_y^2} \right] \\
    &\qquad \times \left[1 -e + \frac{C_x\pi \cos(\pi y/ 2\tilde{L}_y) \cos(\pi x/ \tilde{L}_x)}{\tilde{L}_x} + \frac{A^2 \pi^2 \cos^2(\pi y/ 2 \tilde{L}_y) \sin^2(\pi x / 2 \tilde{L}_x)}{8 \tilde{L}_x^2} \right] \\
    & - \left[ \frac{A^2 \pi^2 \cos(\pi y/2\tilde{L}_y) \cos(\pi x / 2\tilde{L}_x) \sin(\pi x / 2 \tilde{L}_x) \sin(\pi y / 2\tilde{L}_y)}{8 \tilde{L}_x \tilde{L}_y} \right. \\
    &\qquad \left. - \frac{C_y\pi \sin(\pi y / \tilde{L}_y) \sin(\pi x /2\tilde{L}_x)}{4 \tilde{L}_x}- \frac{C_x \pi \sin(\pi y /2 \tilde{L}_y) \sin(\pi x/ \tilde{L}_x)}{4 \tilde{L}_y}\right]^2.
\end{aligned}
\end{equation}

\begin{figure}[!ht]
	\centering
	\includegraphics[width=3.5in]{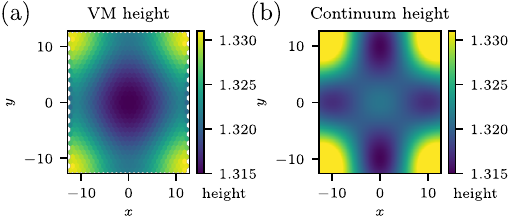}
	\caption{Cell heights for a simply supported, compressed and buckled plate with straight edges, which is the same plate as in Fig.~4(b) in the main manuscript. The plate has size $(27,31)$, tension $\Gamma=0.7$ and compressive stress $\sigma_{xx}=0.03$. (a) VM simulation result for cell heights, shown in undeformed mid-plane coordinates.. (b) Theoretically expected height from continuum approximation.}
\label{fig:buckling_heights}
\end{figure}

Figure~\ref{fig:buckling_heights} depicts the heights of the VM monolayer shown in Fig.~5(b) in the main manuscript, cf.~panel~(a),
and the theoretical height that we expect based on Eq.~(\ref{eq:height_det_buckled}) for the parameters that were found in the minimization, cf.~panel~(b).
The highest order mode in the deformation, i.e.\ the compression (and larger height) in the plate corners, is captured by the theoretical approximated continuum deformation.
However, we do see deviations in the center of the edges and in the center of the plate,
which stem from the neglection of higher order modes in the Fourier representation of the deformation and deflection fields.
Notice, however, that the order of the height changes is comparable and as such the stretching energy should be approximated reasonably well.

\bibliography{references}% Produces the bibliography via BibTeX.